\documentclass[twocolumn,amsmath,amssymb,prl,10pt,nofootinbib,superscriptaddress]{revtex4}
\usepackage{ graphicx, float,amsmath, amsmath, amssymb}
\usepackage[export]{adjustbox}

\def\be{\begin{equation}}
\def\ee{\end{equation}}
\def\bea{\begin{eqnarray}}
\def\eea{\end{eqnarray}}
\def\bse{\begin{subequations}}
\def\ese{\end{subequations}}

\usepackage[breaklinks, colorlinks, citecolor=blue]{hyperref}
\usepackage[labelsep=period]{caption}
\usepackage[normalem]{ulem}
\usepackage{soul}
\usepackage{minitoc}

\usepackage{bm}

\begin{document}
\title{Analytical proof of the isospectrality of quasinormal modes\\ for Schwarzschild-de Sitter and Schwarzschild-Anti de Sitter spacetimes}


\author{Flora Moulin}%
\affiliation{%
Laboratoire de Physique Subatomique et de Cosmologie, Universit\'e Grenoble-Alpes, CNRS/IN2P3\\
53, avenue des Martyrs, 38026 Grenoble cedex, France
}

\author{Aur\'elien Barrau}%
\affiliation{%
Laboratoire de Physique Subatomique et de Cosmologie, Universit\'e Grenoble-Alpes, CNRS/IN2P3\\
53, avenue des Martyrs, 38026 Grenoble cedex, France
}
\affiliation{%
Corresponding author
}

\date{\today}
\begin{abstract} 
The deep reason why the equations describing axial and polar perturbations of Schwarzschild black holes have the same spectrum is far from trivial. In this article, we revisit the original proof and try to make it clearer. Still focusing on uncharged and non-rotating black holes, we extend the results to spacetimes including a cosmological constant, which have so far mostly been investigated numerically from this perspective. 
 \end{abstract}
\maketitle


\section{Introduction}

The direct measurement of gravitational waves emitted by the coalescence of black holes (BHs) is now possible. Since the seminal detection by LIGO \cite{Abbott:2016blz}, several other events were recorded and a catalogue is already available \cite{LIGOScientific:2018mvr}. The recent improvement in sensitivity has even led to a dramatic increase in the detection rate. The recorded gravitational waves carry fundamental informations about the structure of spacetime, BHs being vacuum solutions of the Einstein field equations. Three phases can be distinguished during a coalescence: the inspiral, the merger and the ringdown. The later can be partially treated perturbatively as a superposition of damped oscillations with different complex frequencies, called quasinomal modes (QNMs). An  intuitive introduction can be found in \cite{Chirenti:2017mwe} and a review in \cite{Nollert:1999ji}. The ringdown does not lead to pure ``normal" modes because the system looses energy through the emission of gravitational waves. The equations for the metric perturbations are somehow unusual because of their boundary conditions: the waves have to be purely outgoing at infinity and purely ingoing at the event horizon. The radial part can schematically be written as $\phi \propto e^{-i\omega t} = e^{-i(\omega_R + i\omega_I)t}$ where $\omega_R$ is proportional to the frequency and $\omega_I$ is the inverse of the decaying time scale. The process is stable when $\omega_I<0$. Basically, QNMs are characterized by their overtone and multipole numbers: $n$ and $\ell$.\\

The determination of QNMs have driven a huge amount of efforts (see, {\it e.g.}, \cite{Kokkotas:1999bd} for a historical review, \cite{Berti:2004um,Dorband:2006gg} for an example of quite recent results based on a numerical approach and  \cite{Schutz:1985zz,Iyer:1986np,Iyer:1986nq,Kokkotas:1988fm,Konoplya:2003ii,Konoplya:2019hlu,Moulin:2019ekf} for WKB treatments). This article is not about the calculation of the complex frequencies but about a remarkable -- and quite strange -- property. The perturbations of the metric are described by two different equations depending on their parity: whether polar or axial, they do not fulfill the same equation. They both obey  a Schr\"odinger-like equation (Eq. \ref{waveeq}) but with different potentials. For a spherical time-independent metric, one can write

\bea
ds^2=e^{2 \mu _t} dt^2-e^{2 \psi}(d \phi - q_t dt -q_r dr - q_\theta d \theta )^2 \nonumber \\ - e^{2 \mu _r} dr^2 - e^{2 \mu _\theta} d \theta ^2. 
\eea

For the special case such that

\bea
 \quad e^{2 \mu _t} =e^{- 2 \mu _r}= B(r) , \quad e^{2 \mu _\theta}=r^2, \\ \quad e^{2 \psi} =r^2 sin^2 (\theta) 
\quad \mathrm{and}  \quad q_t = q_r = q_\theta = 0,
\eea

\noindent the perturbations will be described by $q_t$, $ q_r$ and $q_\theta$,  being first 
order small quantities, and $\mu _t$, $\mu  _r$, $\mu _ \theta$, and $\psi$  which receive small increments $\delta \mu _t$, $\delta  \mu _r$, $\delta  \mu _ \theta$ and $\delta  \psi$. The former lead to a non-static stationary distribution of mass-energy leading to a rotating BH. They are called the axial perturbations. The latter do not imply any rotation and are called the polar perturbations. 

In the Schwarzschild case, the Regge-Wheeler potential (for the axial parity) is given by
\begin{equation}
V^{\textrm{RW}}_{\ell}(r) = \left(1-\frac{2M}{r}\right)\left[\frac{\ell(\ell+1)}{r^2} - \frac{6M}{r^3}\right]\,,
\label{eq:RG}
\end{equation}
and the Zerilli potential (for the polar parity) reads
 \begin{eqnarray}
& &V^{\textrm{Z}}_{\ell}(r) = \frac{2}{r^3}\left(1-\frac{2m}{r}\right)\times \nonumber\\
&\times&\frac{9M^3 + 3L^2Mr^2 + L^2(1+L)r^3 + 9M^2Lr}{(3M+Lr)^2}\,,
\label{eq:Ze}
\end{eqnarray}
with $L = \ell(\ell+1)/2 - 1$. The remarquable fact -- known as isospectrality -- is that those equations share the same spectrum of quasinormal modes. This is also true for the Reissner-Nordstr\"om and Kerr metrics. This might appear as a kind of ``miracle" when using the standard tensor formalism where the axial and polar perturbations are teated independently. However, when one actually works in the Newmann-Penrose (NP) formalism \citep{Newman:1961qr}, isospectrality comes as a quite natural feature. This property remains however true only for very specific spacetimes. It is not yet fully clear whether isospectrality is generic or happens as an incredible ``stroke of luck" for classical BHs.\\

In \citep{Bhattacharyya:2018qbe}, it was shown that isospectrality is broken down for general $f(R)$ gravity. In the case of Lovelock black holes, isospectrality is roughly recovered but not exactly \citep{Prasobh2014}.  It fails in Chern-Simons gravity \cite{Bhattacharyya:2018hsj}. The presence of a dilatonic field also breaks isospectrality \cite{Ferrari:2000ep,Brito:2018hjh}. Actually, a perturbative analysis shows that isospectrality seems to be quite generically lost in theories beyond GR \cite{Cardoso:2019mqo}. However, it seems that Schwarzschild-(anti)-de-Sitter (S(A)dS) black holes are isospectral, although the situation is not fully clear \cite{Cardoso:2003cj,Dias:2018ynt,Dias:2018etb,Tattersall:2018axd}.\\

In this article, we try to make clearer the quite involved historical derivation by Chandrasekhar \citep{Chandrasekhar:1985kt} and extend it as far as it can be using the original argument. Although no spectacular new results is obtained, we elegantly end up with an analytical explanation of the isospectrality of SdS and SAdS black holes. We begin with a general metric of the form

\be
ds^2= B(r) dt^2- B(r)^{-1} dr^2-r^2 d \Omega ^2,
\label{metric}
\ee
and explicitly show that if $B(r)$ describes a SdS or SAdS spacetime, the isospectrality property holds. This does not rigorously mean that it is a necessary condition in general but it is one if we rely on the historical strategy to approach isospectrality. The proof can be straightforwardly extended to the case of a charged BHs (the steps are the same than for going from Schwarzschild to Reissner-Nordstr\"om). \\

Our aim here is just to slightly generalize the original derivation and to explain in details each step of the proof. This is mainly useful for pedagogical, methodological and historical purposes. Modern and extremely efficient methods are given in \cite{Glampedakis:2017rar} and \cite{Yurov:2018ynn}. In these references, new results are obtained on the isospectrality, traced back to the fact that the Zerilli and Regge-Wheeler equations are related by a Darboux transformation. More precisely, it is shown that although standard and binary Darboux transformations ensure isospectrality, generalized ones -- associated with long-range potentials -- do not solve exactly the problem. Such methods are powerful and well suited for most complex problems. They also open fascinating mathematical questions that are still unanswered. We however will not use them here and will remain close to the original derivation. The small generalization that we provide is already non-trivial.\\

In the first section, we review sufficient conditions for isospectrality. Then, we introduce the NP formalism which will be used to determine the radial equation. We finally proceed to the full calculation and conclude.

\section{Conditions for isospectrality}

To study  black hole perturbations, we separate the radial and angular parts so as to obtain a wave equation for radial and time variables. This equation has a Schr\"odinger-like form:

\be
\frac{d^2 Z}{dr^{*2}}+ \omega ^2 Z - V Z =0,
\label{waveeq}
\ee

\noindent with $r_*$ the tortoise coordinate defined by $dr_*=dr/B(r)$. The eigenvalue $\omega$ is the frequency of the wave satisfying the boundary conditions given in the introduction and detailed in the following sections.
In full generality, if $Z_2$ satisfies Eq. (\ref{waveeq}) with a potential $V_2$, then  

\be
Z_1 = p Z_2 + q \frac{dZ_2}{dr^*},
\label{c1}
\ee

\noindent  with $p$ and $q$ two functions, also satisfies  Eq. (\ref{waveeq}) with $V_1$ if \cite{Chandrasekhar:1985kt} 

\be 
V_1 = V_2 + \frac{2}{q} \frac{dp}{dr^*} + \frac{1}{q} \frac{d^2 q}{dr^{*2}} ,
\label{c2}
\ee

\noindent and  
\be 
2p\frac{dp}{dr^*} + p \frac{d^2 q}{dr^{*2}} - \frac{d^2 p}{dr^{*2}}q - 2q \frac{d q}{dr^*} (V_2 -\omega ^2) - q^2 \frac{d V_2}{dr^*}=0.
\label{c3}
\ee

We don't use here an index for $\omega$ as the isopectrality precisely means that $\omega_1=\omega_2$. Equation (\ref{c3}) is equivalent to 

\be
p^2+ \bigg( p\frac{dq}{dr^*}- \frac{dp}{dr^*}q \bigg)- q^2 (V_2 - \omega ^2) = C^2 = \mathrm{cte}.
\label{c3eqi}
\ee

To show that Eq. (\ref{c2}) and Eq. (\ref{c3}) imply isospectrality, we use the fact that $Z_2$ satisfies Eq. (\ref{waveeq}), which implies 

\be 
 \frac{d^3 Z_2}{dr^{*3}} + \omega ^2  \frac{d Z_2}{dr^*} -  \frac{d V_2}{dr^*} Z_2- V_2\frac{d Z_2}{dr^*}=0. 
 \label{derivee}
\ee

When replacing $Z_1$ and $V_1$ by their expressions given by Eq. (\ref{c1}) and Eq. (\ref{c2}), we are led to

 \be
 \begin{split}
\frac{d^2 Z_1}{dr^{*2}}+& \omega ^2 Z_1 - V_1 Z_1 = \\
&\bigg( \frac{d^2 p}{dr^{*2}} - \frac{2p}{q}\frac{d p}{dr^*} - \frac{p }{q}\frac{d^2 q}{dr^{*2}} \bigg) Z_2  \ \ \ \ \ \  \ \ \ \ \ \   \ \ \ \ \ \ \ \ \ \ \ \ \ \ \ \\ 
& + q\frac{d V_2}{dr^*} Z_2 + 2\frac{d q}{dr^*}\frac{d^2 Z_2}{dr^{*2}}.
\end{split}
\ee

Using Eq. (\ref{c3}) and Eq. (\ref{derivee}), one can conclude that $Z_1$ satisfies Eq. (\ref{waveeq}) with $V_1$. \\

We first establish the equations governing the gravitational perturbations, we then expose the conditions required to transform it into a wave equation. Finally, we show isospectrality for SdS and SAdS spacetimes by finding the functions $p$ and $q$ satisfying Eq. (\ref{c2}) and Eq. (\ref{c3eqi}) for the potentials of axial and polar perturbations. It should be emphasized that we do not assume a S(A)dS spacetime from the beginning but, instead, are led to it by the requirement that isospectrality emerges -- at least in this approach.

\section{The Newman-Penrose formalism}

To go ahead, the perturbations need to be analyzed in the NP formalism \cite{Newman:1961qr}. This is a special case of the tetrad formalism (see, {\it e.g.}, \cite{deFelice:1990hu}). To guide the unfamiliar reader, we make every step leading to the result explicit in a pedagogical perspective. In this approach, one needs to set up a basis of four null vectors at each point of spacetime. This basis is made of a pair of real null vectors \textbf{l} and \textbf{n} and a pair of complex conjugate null vectors \textbf{m} and $\overline{\textbf{m}}$:

\be
\textbf{l}.\textbf{l}=\textbf{n}.\textbf{n}=\textbf{m}.\textbf{m}=\overline{\textbf{m}}.\overline{\textbf{m}}=0.
\label{pro1}
\ee

Furthermore, these vectors satisfy the following orthogonality relations: 

\be
\textbf{l}.\textbf{m}=\textbf{l}.\overline{\textbf{m}}=\textbf{n}.\textbf{m}=\textbf{n}.\overline{\textbf{m}}=0.
\label{pro2}
\ee

 We also require the normalization

\be
\textbf{l}.\textbf{n}=1 \ \ \mathrm{and} \ \ \textbf{m}.\overline{\textbf{m}}=-1,
\label{pro3}
\ee

\noindent but this latter condition is less crucial in the NP formalism. The number of equations is conveniently reduced thanks to the use of complex numbers.  Any basis with the properties given by Eqs (\ref{pro1}), (\ref{pro2}) and (\ref{pro3}) can be considered. For example, in the Schwarzschild case one usually works with the Kinnersley tetrad and sometimes the Carter one \citep{Batic:2017qcr}. Here, we choose a Kinnersley-like tetrad: 

\bea
l^i= \left( \frac{1}{B(r)}, 1, 0, 0 \right), \\
n^i=\left( \frac{1}{2} , -\frac{B(r)}{2} , 0, 0 \right), \\
m^i= \left( 0, 0, \frac{1}{\sqrt{2}r},\frac{i}{\sqrt{2} r \sin \theta}\right),  \\
\overline{m}^i= \left( 0, 0, \frac{1}{\sqrt{2}r},\frac{-i}{\sqrt{2} sin \theta}\right).
\eea

In the NP formalism, the directional derivatives are usually denoted by the following symbols:

\be 
 D =l^i \partial _i; \ \ \ \   \Delta =n^i \partial _i; \ \ \ \   \delta =m^i \partial _i; \ \   \ \   \delta ^* = \overline{m}^i \partial _i.
 \label{operator}
 \ee

The equations will be written with the so-called spin coefficients \cite{Newman:2009} carrying (roughly speaking) the information on the Riemann tensor. To make things explicit, we switch, here, to the more general framework of the standard tetrad formalism. The four contravariant vectors of the basis are 
$e^i_a,$
 where $a,b,c$ ... are the tetrad indices, indicating the considered vector and $i,j,k$ ... are the tensor indices, indicating the considered componant (alternatively, one can also think to the lower index as an internal Lorentz one and consider the upper index as a coordinate one). The correspondance reads as $\bm{e_1}=\bm{l}$, $\bm{e_2}=\bm{n}$, $\bm{e_3}=\bm{m}$ and $\bm{e_4}=\bm{\overline{m}}$ with $\bm{e^1}=\bm{e_2}$, $\bm{e^2}=\bm{e_1}$, $\bm{e^3}=-\bm{e_4}$ and $\bm{e^4}=-\bm{e_3}$.  For example, $e_{1 2, 3}$ represents the second componant of $\textbf{l}$, derived with respect to $\theta$.  We define the Ricci rotation coefficients (the symbol ``;'' referring to a covariant derivative)

\bea 
\gamma  _{cab} = e_c^k e_{a k;i} e_b^i,
\label{gamma1}
 \eea
 
 \noindent or equivalently
 
 \bea 
 e_{a k;i}= e^c_k  \gamma  _{cab} e^b_i.
 \label{gamma2}
 \eea
 
 These coefficients are antisymmetric with respect to the first pair of indices:
 
 \bea 
\gamma  _{cab} = - \gamma  _{acb} .
\label{gammaantisym}
 \eea

  Let $\textbf{X}$,$\textbf{Y}$and $\textbf{Z}$ be contravariant vector fields: $\textbf{X}$,$\textbf{Y}$, $\textbf{Z} \in T^1_{\ 0}$. The Riemann tensor field $\textbf{R}$ is of type $(1,3)$:
 
 \bea 
 \textbf{R}: T^1_{\ 0} \times T^1_{\ 0} \times T^1_{\ 0} \rightarrow T^1_{\ 0} .
\eea 

\noindent It is defined as 

 \be
 \textbf{R}(\textbf{X},\textbf{Y}) \textbf{Z}=\nabla _{\textbf{X}} \nabla _{\textbf{Y}}\textbf{Z} -\nabla _{\textbf{Y}}\nabla _{\textbf{X}}\textbf{Z},
\ee

\noindent with the  Ricci identity

 \bea 
R^i_{jkl}Z_i=Z_{j;k;l}-Z_{j;l;k}. 
\eea 

\noindent This leads, for $Z=e_a$, to

\bea
R_{ijkl}e_a^i=e_{aj;k;l}-e_{aj;l;k}.
\eea

We project this identity on the tetrad frame and use Eq. (\ref{gamma1}), Eq. (\ref{gamma2}) and Eq. (\ref{gammaantisym}). The projected Riemann tensor depends only on the rotation coefficients and their derivatives:  
 
\begin{align}
R_{abcd} &= R_{ijkl}e_a^i e_b^j e_c^k e_d^l  \nonumber \\
&= [e_{aj; k; l }- e_{aj;l; k} ] e_b^j e_c^k e_d^l  \nonumber \\
&= \bigg( - [ \gamma _{afg} e^f_j e^g_k]_{;l} + [ \gamma _{afg} e^f_j e^g_l ]_{;k} \bigg) e^j_b e^k_c e^l_d \nonumber \\
&= - \gamma  _{abc,d} +\gamma  _{abd,c}  + \gamma  _{baf}(\gamma  _{c \ d}^{\ f}-\gamma  _{d \ c}^{\ f}) \nonumber \\ & +   \gamma  _{fac}\gamma  _{b \ d}^{\ f}-\gamma  _{fad}\gamma  _{b \ c}^{\ f}.
\label{Riemanntetrad}
\end{align}

The spin coefficients of the NP formalism are also defined through the rotation coefficients:

\begin{align*} 
\begin{split}
 \kappa =\gamma  _{311}, \quad  \rho = \gamma  _{314},  \quad   \epsilon =\frac{1}{2}( \gamma  _{211} +  \gamma  _{341} ), \\  
 \sigma = \gamma  _{313}, \quad   \mu = \gamma  _{243},  \quad   \gamma = \frac{1}{2}( \gamma  _{212} +  \gamma  _{342} ) , \\
  \lambda = \gamma  _{244},  \quad    \tau = \gamma  _{312},  \quad   \alpha = \frac{1}{2}( \gamma  _{214} +  \gamma  _{344} ),  \\
\nu = \gamma  _{242} ,\quad   \pi = \gamma  _{241},  \quad   \beta = \frac{1}{2}( \gamma  _{213} +  \gamma  _{343} ). 
\label{spincoef}
\end{split}
\end{align*}

The 36 equations  (\ref{Riemanntetrad}) can be written as 18 complex equations. The 10 independent components of the Weyl tensor $C_{abcd}$ are represented by five complex scalars:

\be
  \begin{split}
  \Psi _0= - C_{1313} &= - C_{pqrs} l^p m^q l^r m^s, \\
    \Psi _1= - C_{1213} &= - C_{pqrs} l^p n^q l^r m^s, \\
      \Psi _2= - C_{1342} &= - C_{pqrs} l^p m^q \overline{m}^r n^s, \\
        \Psi _3= - C_{1242} &= - C_{pqrs} l^p n^q \overline{m}^r n^s, \\
          \Psi _4= - C_{2424} &= - C_{pqrs} n^p \overline{m}^q n^r \overline{m}^s,
  \end{split}
\ee

\noindent and the 20 linearly independent Bianchi identities can be written as eight complex and four real equations. As it will be useful later we also define the following scalars:

\begin{align}
 &\Phi _{00} = -\tfrac{1}{2} R_{11} ; \quad  \Phi _{22} = -\tfrac{1}{2} R_{22} ; \quad  \Phi _{02} = -\tfrac{1}{2} R_{33} ;  \nonumber \\
&  \Phi _{20} = -\tfrac{1}{2} R_{44} ;  \quad  \Phi _{11} = -\tfrac{1}{4}( R_{12}+  R_{34}); \quad  \Phi _{01} = -\tfrac{1}{2} R_{13} ;
   \nonumber \\
&  \Phi _{10} = -\tfrac{1}{2} R_{14} ; \quad  \Phi _{12} = -\tfrac{1}{2} R_{23} ; \quad  \Phi _{21} = -\tfrac{1}{2} R_{24}.
\end{align}


\section{Preliminaries on isospectrality}

\subsection{Derivation of the radial equation}

We assume that the perturbations have a $t$ and $\phi$ dependance given by
$e^{i(\omega t+m \phi )}$ and we define the following operators ($n$ being an integer):

\be
\mathcal{D} _n = \partial _r + \frac{i \omega}{B(r)} + n \bigg( \frac{B'(r)}{B(r)} + \frac{2}{r} \bigg),
\ee
and
\be
\mathcal{L} _n = \partial _{\theta} + \frac{m}{\sin (\theta)} + n.\cot (\theta).
\ee

The prime denotes the derivative with respect to r. Let $\mathcal{D} _n^{\dag}$ be the complex conjugate of $\mathcal{D} _n$ and $\mathcal{L} _n^{\dag} (\theta) =-\mathcal{L} _n (\pi -\theta)$. It is interesting to notice that 
\be 
Br^2\mathcal{D} _{n+1}= \mathcal{D} _n Br^2.
\label{propertyD}
\ee

\noindent The directional derivative given by Eq. (\ref{operator}) reads 
\be 
  \begin{split} 
D= \mathcal{D} _0 , \quad   \quad 
\Delta = -\frac{B(r)}{2}\mathcal{D} _0^{\dag}, \\
\delta= \frac{1}{\sqrt{2}r}\mathcal{L} _0 ^{\dag},
  \quad   \quad 
\delta ^*= \frac{1}{\sqrt{2}r}\mathcal{L} _0.
  \end{split}
\ee

\noindent The five scalars are:

\be 
\Psi _2=  \frac{-2+ 2B(r) - 2 r B'(r) + r^2 B''(r)}{12r^2},
\label{psi2}
\ee

\be
 \Psi _0=\Psi _1=\Psi _3=\Psi _4=0.
 \label{psinul}
\ee

 As $\Psi _0$, $\Psi _1$, $\Psi _3$ and $\Psi _4$ vanish but $\Psi _2$ doesn't, the spacetime defined by Eq. (\ref{metric}) is a Petrov type D spacetime. A corollary of the Goldberg-Sachs theorem \citep{CasalsiCasanellas:2008xt} shows that this implies that $ \kappa $, $\sigma $, $\lambda $, and $\nu $ do vanish. The explicit calculation indeed leads to:
 
  \bea
  \kappa = \sigma = \lambda = \nu  = 0 , 
\label{coefnul}  
  \eea  
  
    \bea
 \tau = \pi = \epsilon = 0 
\label{coefnul}  
  \eea   
 
 \noindent and 
  \bea 
 &\rho =-\frac{1}{r} , \ \  \   \mu =-\frac{B}{2r}  , \  \  \  \gamma =\frac{B'}{4}, \nonumber  \\
   \\
  &\alpha = -\beta = - \frac{\cot \theta}{2 \sqrt{2}r}.\nonumber
 \label{coef}
\eea

There are 6 linearized equations, 2 from the Ricci identities (\ref{Ricci1},\ref{Ricci2}) and 4 from the Bianchi identities (\ref{Bianchi1},\ref{Bianchi2},\ref{Bianchi3}, \ref{Bianchi4}):

\begin{align}
&( \delta ^* -4 \alpha ) \Psi _0 - (D  - 4 \rho ) \Psi _1 = 3 \kappa \Psi _2 + [R_1], \label{Bianchi1}\\
&( \Delta - 4 \gamma + \mu ) \Psi _0 - (\delta  + 2 \alpha ) \Psi _1 = 3 \sigma \Psi _2 + [R_2], \label{Bianchi2}\\
&( D - 2\rho ) \sigma - (\delta +2 \alpha ) \kappa = \Psi _0, \label{Ricci1}
\end{align}

\noindent and

\begin{align}
&( D  - \rho ) \Psi _4 - (\delta ^*  + 2 \alpha ) \Psi _3 = - 3 \lambda \Psi _2 + [R_3], \label{Bianchi3}\\
&( \delta - 4 \alpha  ) \Psi _4 - ( \Delta + 2 \gamma + 4 \mu ) \Psi _3 = - 3 \nu \Psi _2  + [R_4], \label{Bianchi4}\\
&( \Delta + 2\mu  + 2 \gamma  ) \lambda - (\delta ^* + 2 \alpha  + ) \nu = - \Psi _4, \label{Ricci2}
\end{align}
with 

\begin{align}
[R_1]=& - D \Phi _{01} + \delta \Phi _{00} + 2 \rho \Phi _{01} \nonumber \\ &+ 2 \sigma \Phi _{10} -2 \kappa \Phi_{11}- \kappa \Phi _{02} \nonumber \\
=& \tfrac{\kappa}{4 r^2} [ 2 - 2 B(r) +r^2 B''(r) ] \\
[R_2] =&  -D \Phi_{02}+ \delta \Phi_{01}+ 2 \alpha  \Phi_{01}\nonumber \\ & - 2 \kappa \Phi _{12} - \lambda \Phi _{00} + 2 \sigma \Phi _{11} + \rho \Phi _{02}  \nonumber \\
=& \tfrac{- \sigma}{4 r^2} [ 2 - 2 B(r) +r^2 B''(r) ] \\
[R_3] =& - \Delta  \Phi _{02} + \delta ^* \Phi _{21} + 2 \alpha \Phi _{21} \nonumber \\ & + 2 \nu \Phi _{10} + \sigma \Phi _{22} - 2 \lambda \Phi _{11} - \mu \Phi _{20}  \nonumber \\
=& \tfrac{\lambda}{4 r^2} [ 2 - 2 B(r) +r^2 B''(r) ] \\
[R_4] =& \Delta \Phi _{21} - \delta ^* \Phi _{22} + 2( \mu + \gamma) \Phi _{21} \nonumber \\ &- 2 \nu \Phi _{11} - \nu \Phi _{20} + 2 \lambda \Phi _{12}  \nonumber \\
=& \tfrac{\nu}{4 r^2} [ 2 - 2 B(r) +r^2 B''(r)], 
\end{align}

\noindent where $\Psi _0$, $\Psi _1$, $\Psi _3$, $\Psi _4$, $\kappa$, $\sigma$, $\lambda$, and $\nu$ are the perturbations. Using Eq. (\ref{psi2}), Eq.(\ref{coefnul}), and Eq. (\ref{coef}), we obtain:

\begin{align}
&\frac{1}{r \sqrt{2}}  \bigg( \mathcal{L}_0 +2 \cot \theta \bigg) \Psi _0- \bigg( \mathcal{D} _0 + \frac{4}{r} \bigg) \Psi _1 =  3 \kappa \Psi _2 +[R_1], \nonumber  \\  &  \\ 
&-\frac{B}{2} \bigg( \mathcal{D}_0 ^{\dag} + \frac{2 B'}{B} + \frac{1}{r}  \bigg) \Psi _0 -  \frac{1}{r \sqrt{2}}  \bigg( \mathcal{L}_0^{\dag} - \cot \theta \bigg) \Psi _1  \nonumber  \\& =   3 \sigma  \Psi _2 +[R_2],\\
 &\bigg( \mathcal{D} _0 + \frac{2}{r} \bigg) \sigma  - \frac{1}{r \sqrt{2}}  \bigg( \mathcal{L}_0^{\dag} - \cot \theta \bigg) \kappa   =  \Psi _0, \\
& \bigg( \mathcal{D} _0 + \frac{1}{r} \bigg) \Psi _4 - \frac{1}{r \sqrt{2}}  \bigg( \mathcal{L}_0 - \cot \theta \bigg) \Psi _3 =  -  3 \Psi _2 \lambda +[R_3], \\
 &\frac{1}{r \sqrt{2}}  \bigg( \mathcal{L}_0^{\dag}  +2 \cot \theta \bigg) \Psi _4 +  \frac{B}{2} \bigg( \mathcal{D}_0 ^{\dag} - \frac{ B'}{B} + \frac{4}{r}  \bigg) \Psi _3  \nonumber  \\& =   - 3 \Psi _2  \nu +[R_4], \\
& -\frac{B}{2} \bigg( \mathcal{D}_0 ^{\dag} - \frac{ B'}{B} + \frac{2}{r}  \bigg) \lambda - \frac{1}{r \sqrt{2}}  \bigg( \mathcal{L}_0 - \cot \theta \bigg) \nu  =    \Psi _4.
\end{align}

\noindent We proceed to the following change of variables:

\bea
\Phi _0= \Psi _0, \ \   \Phi _1= \Psi _1 r \sqrt{2}, \ \  k= \frac{\kappa}{r^2 \sqrt{2}}, \ \  s=\frac{\sigma}{r}, \\
\Phi _4= \Psi _4 r^4, \ \  \Phi _3= \Psi _3 \frac{r^3}{\sqrt{2}}, \ \  l= \frac{\lambda r}{2}, \ \   n=\frac{\nu r^2}{\sqrt{2}}.
\eea

\noindent This leads to:

\begin{align}
 &\mathcal{L}_2  \Phi _0 - \bigg( \mathcal{D} _0 + \frac{3}{r} \bigg) \Phi _1= 6 r^3 \Psi _2 k + \sqrt{2} r [R_1],   \label{one} \\ 
&B r^2 \bigg( \mathcal{D}_2 ^{\dag} - \frac{3}{r}  \bigg) \Phi _0 +  \mathcal{L}_{-1}^{\dag} \Phi _1 = - 6 r^3 \Psi _2 s - 2 r^2 [R_2] , \label{two} \\
&\bigg( \mathcal{D} _0 + \frac{3}{r} \bigg) s  - \mathcal{L}_{-1}  k = \frac{\Phi _0}{r}, 
\label{three} \\
&\bigg( \mathcal{D} _0 - \frac{3}{r} \bigg) \Phi _4 -  \mathcal{L}_{-1} \Phi _3 = - 6 r^3 \Psi _2 l +  r^4 [R_3], \label{four}\\
& \mathcal{L}_2^{\dag}  \Phi _4 +\frac{Br^2}{2} \bigg( \mathcal{D}_{-1} ^{\dag} + \frac{3}{r}  \bigg) \Phi _3 = -6 r^3 \Psi _2 n + \sqrt{2} r^5 [R_4],\label{five}\\
&Br^2 \bigg( \mathcal{D}_{-1} ^{\dag} + \frac{3}{r}  \bigg) l + \mathcal{L}_{-1} n= \frac{\Phi _4}{r}.  \label{six}
\end{align}

By applying $\mathcal{L}_{-1}^{\dag}$ to Eq. (\ref{one}) and $\bigg( \mathcal{D} _0 + \tfrac{3}{r} \bigg)$ to Eq. (\ref{two}), we are then led to:

\begin{align}
&\mathcal{L}_{-1}^{\dag} \mathcal{L}_2 \Phi _0 + \bigg( \mathcal{D}_0 + \frac{3}{r} \bigg) \bigg[ Br^2 \bigg( \mathcal{D}_2^{\dag} - \frac{3}{r} \bigg) \Phi _0 \bigg] =  \nonumber \\ 
& 6 r^3 \Psi _2  \bigg[ \mathcal{L}_{-1}^{\dag}k - \bigg(\mathcal{D}_0 + \frac{3}{r} \bigg) s \bigg] - 6 s \partial _r (r^3 \Psi _2 ) \nonumber \\ &+ \mathcal{L}_{-1}^{\dag} (\sqrt{2}r[R_1]) - 2 \bigg(\mathcal{D}_0 + \frac{3}{r} \bigg)(r^2 [R_2]).
\end{align}

It should be noticed that if  $r^3 \Psi _2$  is a constant and if $[R_1]$ and $[R_2]$ do vanish, then the left part of Eq. (\ref{three}) does appear and can be replaced by $\Phi _0/r$ which leads to a decoupled equation for $\Phi _0$:

\begin{align}
&\mathcal{L}_{-1}^{\dag} \mathcal{L}_2 \Phi _0 + \bigg( \mathcal{D}_0 + \frac{3}{r} \bigg) \bigg[ Br^2 \bigg( \mathcal{D}_2^{\dag} - \frac{3}{r} \bigg) \Phi _0  \bigg] = -  6 r^2 \Psi _2  \Phi _0.
\label{first}  
\end{align}

 In the same way,  by applying $\mathcal{L}_{-1}$ to Eq. (\ref{five}) and $B r^2 \bigg( \mathcal{D} _{-1}^{\dag} + \tfrac{3}{r} \bigg)$ to Eq. (\ref{four}), we can obtain a decoupled equation for $ \Phi _4$ if, in addition,  $[R_3]$ and $[R_4]$ are zero:

\begin{align}
Br^2 \bigg( \mathcal{D} _{-1}^{\dag} + \tfrac{3}{r} \bigg) \bigg[ \bigg( \mathcal{D} _0 - \tfrac{3}{r} \bigg) \Phi _4  \bigg]+ \mathcal{L}_{-1} \mathcal{L}_2^{\dag} \Phi _4 = -  6 r^2 \Psi _2  \Phi _4, \label{second}   
\end{align}

\noindent where Eq. (\ref{six}) has also been used. To summarize, one is led to two decoupled equations, for $\Phi_0$ and $\Phi_4$, if:

\be 
r^3 \Psi _2= \mathrm{cte} \label{separ1}
\ee

\noindent and 

\be  
[R_1], [R_2], [R_3], [R_4]= 0. \label{separ2}
\ee

The first condition, Eq. (\ref{separ1}), implies that the metric must have the form

\be 
B(r) = 1 +\frac{C_1}{r} + C_2 r + C_3 r^2,
\ee
while the second condition, Eq. (\ref{separ2}), implies 

\be 
B(r) = 1 +\frac{D_1}{r} + D_2 r^2,
\ee

\noindent with $C_i $ and $ D_i $ some arbitrary constants. The latter condition (which contains the previous one) corresponds to Schwarzschild-de Sitter and Schwarzschild-Anti-de Sitter spacetimes: 

\be 
B(r)= 1 - \frac{2M}{r} - \Lambda r^2,
\label{metric}
\ee

\noindent with $\Lambda$ the cosmological constant. \\

In the Reissner-Nordstr\"om case, it is possible to find new variables that mix the $\Phi _i$ functions with the spin coefficients so that a separation is possible \citep{Chandrasekhar:1985kt}. This works because $\Psi _1$ does not vanish and this implies two more equations which lead to a radial equation of the form of Eq. (\ref{equation1}) such that $P$ and $Q$ lead to isospectrality. As far as our argument is concerned, the extension from the Schwarzschild case to the charged case is therefore straightforward.

\section{Proof of isospectrality}

The equations (\ref{first}) and (\ref{second}) read

\bea
\bigg[ \mathcal{L}_{-1} ^{\dag} \mathcal{L}_2 + Br^2\mathcal{D}_1 \mathcal{D}_2 ^{\dag} -6 (\Lambda r + i \omega)r \bigg] \Phi _0=0, \\
\bigg[ \mathcal{L}_{-1}  \mathcal{L}_2^{\dag} + Br^2\mathcal{D}_{-1}^{\dag} \mathcal{D}_0  -6 (\Lambda r - i \omega)r \bigg] \Phi _4=0.
\eea

\noindent If we set 
\bea
\Phi _0 = R_{+2}(r) S_{+2}(\theta), \ \ \  \Phi _4 = R_{-2}(r) S_{-2}(\theta),
\eea

\noindent they are separable with a separation constant $\mu ^2$. This leads to:

\bea
\mathcal{L}_{-1} ^{\dag} \mathcal{L}_2 S_{+2} = -\mu ^2 S_{+2}, \label{248}\\
\bigg[ Br^2\mathcal{D}_1 \mathcal{D}_2 ^{\dag} -6 (\Lambda r  + i \omega)r \bigg] R_{+2} = \mu ^2 R_{+2}, \label{249}\\
\mathcal{L}_{-1}  \mathcal{L}_2^{\dag} S_{-2} = - \mu ^2 S_{-2}, \label{250}\\
\bigg[ Br^2\mathcal{D}_{-1}^{\dag} \mathcal{D}_0  -6 (\Lambda r + i \omega)r \bigg] R_{-2} = \mu ^2 R_{-2}.
\label{251}
\eea

The separation constant is calculated with Eq. (\ref{248})  -- or Eq. (\ref{250}) -- by requiring the regularity of $S_{+2}$ at $\theta=0$ and $\theta= \pi$. The angular equation is the same than in the Schwarzschild case, which gives $\mu ^2= l(l+1)-2=2L$. \\

We set 

\be 
\mathcal{D}_0 = \frac{1}{B} \Lambda _+,  \quad  \quad \mathcal{D}_0^{\dag} = \frac{1}{B} \Lambda _-.
\ee
Using the the tortoise coordinate $r*$ (with $ \frac{d}{dr^*}=B \frac{d}{dr}$), we are led to

\be 
\Lambda _+ = \frac{d}{dr^*} + i \omega ,  \quad \quad
\Lambda _- = \frac{d}{dr^*} - i \omega \quad  \mathrm{and} \quad \Lambda ^2 = \Lambda _+ \Lambda _- ,
\ee

\noindent that is
\be
\Lambda _{\pm} = \Lambda _{\mp} \pm 2i \omega.
\ee

The operator $\Lambda ^2 $ has no link with the cosmological constant (and cannot be confused with it as the cosmological constant never appears squared in this article).  It should be pointed that the equation

\be
\bigg[ Br^2\mathcal{D}_{-1} \mathcal{D}_0 ^{\dag} -6 (\Lambda r + i \omega)r \bigg] B^2 r^4 R_{+2} = \mu ^2 B^2 r^4 R_{+2}
\label{257}
\ee

\noindent is the same than Eq. (\ref{249}). Using the properties of Eq. (\ref{propertyD}), we obtain

\be
Br^2\mathcal{D}_{-1} \mathcal{D}_0 ^{\dag}= (Br^2)^2\mathcal{D}_0 \frac{1}{Br^2} \mathcal{D}_0 ^{\dag} = r^4 B \Lambda _+ \bigg( \frac{1}{B^2 r^2} \Lambda _-\bigg).
\ee

Defining $Y$ as

\be
 Y=B^2r R_{+2},
\ee

\noindent we are led to 
\be
\begin{split}
\Lambda _+ \bigg( \frac{1}{Br^2} \Lambda _- (r^3 Y) \bigg)  = \\   \frac{r}{B^2} \Lambda ^2 Y+ \frac{d}{dr^*} \bigg(\frac{r}{B^2}\bigg) \Lambda _- Y + \frac{3}{B} \Lambda _+ Y + \frac{d}{dr^*} \bigg(\frac{3}{B}\bigg) Y.
\label{calcul}
\end{split}
\ee

By calculating the derivative and replacing $\Lambda _+$ by $\Lambda _- + 2i \omega $ in Eq. (\ref{calcul}), we find that Eq. (\ref{257}) is equivalent to

\be
\Lambda ^2 Y + P \Lambda _- Y - Q Y = 0,
\label{equation1}
\ee

\noindent with 

\be
\begin{split}
P&=\bigg( \frac{4B}{r} - 2B'  \bigg) \\
&= \frac{B^2}{r^4} \frac{d}{dr^*} \bigg( \frac{r^4}{B^2}  \bigg) \\
&= \frac{d}{dr^*} \bigg( \log \bigg( \frac{r^4}{B^2} \bigg) \bigg),
\label{Plog}
\end{split}
\ee
\noindent and
\be
Q=\bigg( \frac{3}{r}B B' + 6 B \Lambda + \mu ^2 \frac{B}{r^2} \bigg) .
\ee
For the same reasons, $Y_{-2}= r^{-3} R_{-2}$, satisfies

\be
\Lambda ^2 Y_{-2} + P \Lambda _+ Y_{-2} - Q Y_{-2} = 0.
\ee

Equation (\ref{equation1}) needs to be transformed into a wave equation in one dimension: 

\be
\Lambda ^2 Z= VZ. 
\label{waveeq1}
\ee

The functions $Y$ and $Z$ both satisfying a second order equation, we write Y as a linear combination of Z and its derivative: 

\bea
Y &=&\zeta \Lambda _+ \Lambda _+  Z + W \Lambda _+ Z \nonumber  \\
&=& \zeta V Z + (W + 2i \omega \zeta ) \Lambda _+ Z,
\label{Y2}
\eea

\noindent with $\zeta$ and $W$ two functions of $r^*$. Applying $\Lambda _-$ to Eq. (\ref{Y2}) yields 
\bea
\Lambda _- Y
&=& \bigg[ \frac{d}{dr^*}(\zeta V ) + WV \bigg]Z  \nonumber  \\ 
&+& \bigg[ \zeta V + \frac{d}{dr^*} (W +2i \omega \zeta ) \bigg] \Lambda _+ Z \nonumber  \\ 
&=& - \gamma \frac{B^2}{r^4} Z + R \Lambda _+ Z,
\label{292}
\eea

\noindent  with 

\bea
R &=& \zeta V + \frac{d}{dr^*} (W + 2 i \omega \zeta ),  \label{291}\\
\gamma &=& - \frac{r^4}{B^2} \bigg( \frac{d}{dr^*} ( \zeta V ) + WV \bigg) .  \label{290}
\eea

 By applying again $\Lambda _- $ to Eq. (\ref{292}), we obtain

\be
\begin{split}
\Lambda _- \Lambda _- Y = \bigg[ - \gamma \frac{B^2}{r^4} + \frac{dR}{dr^*} \bigg] \Lambda _+ Z  \\  
 + \bigg[ 2i \omega  \gamma  \frac{B^2}{r^4} - \frac{d \gamma}{dr^*} \frac{B^2}{r^4} - \gamma \frac{d}{dr*} \bigg( \frac{B^2}{r^4} \bigg) + RV \bigg] Z.
 \label{293}
 \end{split}
\ee

 On the other hand, one can notice that Eq. (\ref{equation1}) leads to: 

\be
\begin{split}
\Lambda _- \Lambda _- Y = -(P +2i \omega) \Lambda _- Y + Q Y \\
= \bigg[ - (P +2i \omega) R + Q(W+2i \omega \zeta) \bigg] \Lambda _+ Z \\ 
+ \bigg[( P+2i \omega) \frac{\gamma B^2}{r^4} + Q \zeta V \bigg] Z.
 \label{295}
 \end{split}
\ee

 Identifying Eq.  (\ref{293}) and Eq. (\ref{295}), and by using the definition of $P$ given by Eq. (\ref{Plog}), we find:

\be
\begin{split}
& - \frac{d \gamma}{dr^*} \frac{B^2}{r^4} - \gamma \frac{d}{dr*} \bigg( \frac{B^2}{r^4} \bigg) + RV \\ &= 
 \frac{d}{dr^*} \bigg( \log \bigg( \frac{r^4}{B^2} \bigg) \bigg) \frac{\gamma B^2}{r^4} + Q \zeta V \\
 &=  - \gamma\frac{d}{dr^*} \bigg( \frac{B^2}{r^4}  \bigg)  + Q \zeta V, \\
 \end{split}
\ee

\noindent which gives

\be
- \frac{d \gamma}{dr^*} \frac{B^2}{r^4}  \\ = 
 (Q \zeta -R )V,
 \label{298}
\ee

\noindent  and

\be
\begin{split}
 \frac{dR}{dr^*} -\frac{B^2}{r^4}  \gamma =   Q(W+2i \omega \zeta)- (P +2i \omega)  R,
  \end{split}
 \ee

\be
\begin{split}
 \frac{r^4}{B^2} \frac{dR}{dr^*} +  \frac{r^4}{B^2}\frac{d}{dr^*} \bigg( \log \bigg( \frac{r^4}{B^2} \bigg) \bigg)R = \\ \gamma + \frac{r^4}{B^2} Q(W+2i \omega \zeta) - 2i \omega  \frac{r^4}{B^2}R,
  \end{split}
 \ee

\be
\begin{split}
\frac{d}{dr^*} \bigg(  \frac{r^4}{B^2} R \bigg) =  \gamma + \frac{r^4}{B^2} \bigg( Q(W+2i \omega \zeta) - 2i \omega R \bigg).
 \label{299}
  \end{split}
 \ee

The combination $\zeta V \times$ Eq. (\ref{299}) + $R \times$ Eq. (\ref{290}) - $\gamma \times$ Eq. (\ref{291}) - $\frac{r^4}{B^2}(W+2i \omega \zeta) \times $ Eq. (\ref{298}) leads to  

\be
\begin{split}
\zeta V \frac{d}{dr^*} \bigg(  \frac{r^4}{B^2} R \bigg) +   \frac{r^4}{B^2} R \frac{d(\zeta V )}{dr^*}  +\\  \gamma \frac{d}{dr^*} (W+2i \omega \zeta) + (W+2i \omega \zeta) \frac{d \gamma}{dr^*}=0, 
 \end{split}
 \ee

\noindent that is to say

\be
\begin{split}
\frac{r^4}{B^2} R \zeta V   + \gamma  (W+2i \omega \zeta) = K = \mathrm{cte}. 
\label{constte}
 \end{split}
 \ee

As we have written $Y$ as a linear combination of $Z$ and $\Lambda _+ Z$ in Eq. (\ref{Y2}), it is possible to write $Z$ as a linear combination of $Y$ and $\Lambda _+ Y$. Using Eq. (\ref{Y2}) and Eq. (\ref{292}):

\be
\begin{split}
KZ &=  \frac{r^4}{B^2} R \zeta V Z  + \gamma  (W+2i \omega \zeta)Z \\
&=  \frac{r^4}{B^2} R Y - \frac{r^4}{B^2} (W+2i \omega \zeta) (\Lambda _- Y + \gamma \frac{B^2}{r^4} Z )\\
&+ \gamma  (W+2i \omega \zeta)Z \\
&= \frac{r^4}{B^2} R Y - \frac{r^4}{B^2} (W+2i \omega \zeta) \Lambda _- Y ,
\end{split}
 \ee

\noindent and 

\be
\begin{split}
K\Lambda _+ Z &=  \frac{r^4}{B^2} R \zeta V \Lambda _+ Z  + \gamma  (W+2i \omega \zeta)\Lambda _+ Z \\
&=\frac{r^4}{B^2}  \zeta V \Lambda _- Y + \frac{r^4}{B^2} \zeta V \gamma \frac{B^2}{r^4} Z + \gamma Y -\gamma \zeta V Z \\
&= \frac{r^4}{B^2}  \zeta V \Lambda _- Y + \gamma Y.
\end{split}
 \ee
 
 By requiring $\gamma = \mathrm{cte}$ and $ \zeta=1$, Eq. (\ref{298}) leads to

\be 
R=Q,
\ee

\noindent and from Eq. (\ref{291}) one obtains

\be 
V= Q - \frac{dW}{dr^*}.
\ee

\noindent  Equation (\ref{299}) then leads to

\be 
\frac{d}{dr^*} \bigg(  \frac{r^4}{B^2} R \bigg) =  \gamma + \frac{r^4}{B^2} QW,
\label{305}
\ee

\noindent and Eq. (\ref{constte}) yields

\be
\frac{r^4}{B^2} Q V + \gamma W = K-2i \omega \gamma = \kappa = \mathrm{cte}.
\label{306}
\ee

\noindent  Defining 

\be 
F \equiv \frac{r^4}{B^2} Q, 
\label{defF}
\ee

\noindent Eq. (\ref{305}) and Eq. (\ref{306}) lead to

\be 
W= \frac{1}{F} \bigg( \frac{dF}{dr^*} -\gamma \bigg),
\ee

\noindent  and 

\be 
FV + \gamma W = F \bigg( Q- \frac{dW}{dr^*} \bigg)+ \gamma W = \kappa,
\ee

\be
 FQ - F \frac{d}{dr^*} \bigg[ \frac{1}{F} \frac{dF}{dr^*}-\frac{\gamma}{F} \bigg] + \frac{\gamma}{F} \bigg( \frac{dF}{dr^*}-\gamma \bigg)= \kappa,
\ee

\noindent which gives

\be 
\frac{1}{F} \bigg(\frac{dF}{dr^*} \bigg)^2 - \frac{d^2F}{dr^{*2}} + \frac{B^2}{r^4}F^2= \frac{\gamma ^2}{F} + \kappa.
\label{311}
\ee

There exist constants $\gamma$ and $\kappa$ such that Eq. (\ref{311}) is satisfied by the function (\ref{defF}). Depending on the square root of
$\gamma ^2$ chosen ($- \gamma $ or $+ \gamma$), one is led to the equation for  axial or polar perturbations. With

\be 
W^{\pm} = \frac{1}{F} \bigg( \frac{dF}{dr^*}\mp \gamma \bigg),
\ee

\noindent then 

\be 
V^{\pm} = Q -  \frac{d}{dr^*}\bigg(  \frac{1}{F} \frac{dF}{dr}\mp \gamma \bigg). 
\ee

\noindent Defining $f \equiv \frac{1}{F} $,

\be 
V^{\pm} = \pm \gamma \frac{df}{dr^*} + \gamma ^2 f^2 + \kappa f, 
\label{pote}
\ee

\noindent and

\bea
Y &=& V^{\pm}Z^{\pm}+(W^{\pm}+2i \omega) \Lambda _+  Z^{\pm}, \label{318-1}\\
\Lambda _- Y &=& \mp \gamma \frac{B^2}{r^4} Z^{\pm} + Q \Lambda _+ Z^{\pm}, \label{318-2}\\
K^{\pm} &=& \kappa \pm 2i \omega \gamma, \\
K^{\pm} Z^{\pm} &=&  \frac{r^4}{B^2} [ QY -(W^{\pm} +2i \omega) \Lambda _- Y]  \label{319-1},\\
K^{\pm} \Lambda _+ Z^{\pm} &=& \frac{r^4}{B^2}V^{\pm} \Lambda _- Y \pm  \gamma Y.\label{319-2}
\eea

By inserting eq. (\ref{318-1}) and Eq. (\ref{318-2}) in Eq. (\ref{319-1}), one obtains

\be
\begin{split}
K^-  Z^- &=  \frac{r^4}{B^2} \bigg[Q[V^+ Z^+ +(W^+  + 2i \omega) \Lambda _+ Z^+ ]\\ & -(W^- +2i \omega ) [-\gamma \frac{B^2}{r^4} Z^+ + Q \Lambda _+ Z^+] \bigg] \\
&= \bigg[ \frac{r^4}{B^2}Q V^+ + \gamma (W^+ +2i \omega ) - \gamma (W^+ - W^-) \bigg] Z^+ \\
& + F [W^+ - W^-] \Lambda _+ Z^+,
\end{split}
 \ee

\noindent which simplifies to 

 \be 
( \kappa -2i \omega \gamma ) Z^- = (\kappa + 2 \gamma ^2 f) Z^+ -2 \gamma \frac{dZ^+}{dr^*}.
\ee

Equivalently, one can show that
 
\be 
 ( \kappa +2i \omega \gamma ) Z^+ = (\kappa + 2 \gamma ^2 f) Z^- +2 \gamma \frac{dZ^-}{dr^*}.
 \label{e150}
\ee
 
By identification with the previously given condition we are led to 

\be 
q=2 \gamma \  \  \mathrm{and}  \  \  p= \kappa + 2 \gamma ^2 f.
\label{pq}
\ee

Conditions given by Eq. (\ref{c1}), Eq. (\ref{c2}) and Eq. (\ref{c3}) are therefore respected. In \cite{Chandrasekhar:1985kt}, it is shown that if $\omega$ is a characteristic frequency and $Z^- (\omega)$ is a solution belonging to it, then the solution $Z^+ (\omega)$ in accordance with the relation (\ref{e150}), will satisfy the boundary conditions of the quasi normal modes :

\begin{align}
Z^{\pm} &\rightarrow A^{\pm} (\omega )  e^{-i \omega r} \quad (r_* \rightarrow + \infty ) \\
&\rightarrow \quad \quad \quad  e^{+i \omega r_*} \quad (r_* \rightarrow - \infty )
\end{align}

 \noindent  with 

\begin{equation}
A^+ (\omega ) = A^- (\omega ) \frac{\kappa -2 i \omega \gamma}{\kappa +2 i \omega \gamma}.
\end{equation}
  

The values of $\kappa$ and $\gamma$ when the metric function $B(r)$ is defined by (\ref{metric}) now need to be determined. First, one can notice that:

\be 
-F \bigg( \frac{d^2}{dr^{*2}}\log F \bigg)= \frac{1}{F} \bigg(\frac{dF}{dr^*} \bigg)^2 - \frac{d^2F}{dr^{*2}},
\ee

\noindent which implies that Eq. (\ref{311}) reads as 

\be 
-F \bigg[ \frac{d^2}{dr^{*2}} \bigg( \log F \bigg) -\frac{B^2}{r^4}F \bigg]= \frac{\gamma ^2}{F} + \kappa. 
\label{bigone}
\ee

\noindent Moreover, $F$ is given by 

\be
F = \frac{r}{B}( \mu ^2 r + 6M).
\ee

\noindent This leads to
\begin{align}
\frac{B}{F} \frac{dF}{dr}= \frac{\mu ^2 r}{F} + \frac{1}{r}- \frac{4M}{r^2} + \Lambda r,
\end{align}
and
\be
\begin{split}
& \frac{d}{dr}\bigg( \frac{B}{F} \frac{dF}{dr} \bigg) \\ &= -\frac{1}{r^2} + \frac{8M}{r^3}+ \Lambda + \frac{\mu ^2 }{F} \bigg( 1-  \frac{r}{F}\frac{dF}{dr} \bigg)  \\ 
&= -\frac{1}{r^2}  + \frac{8M}{r^3} +    \Lambda + \frac{\mu ^2 }{F B r^2} \bigg(2Mr - 2 \Lambda r^4 - \frac{\mu ^2r^4}{F} \bigg),
\end{split}
\ee
together with
\be
\begin{split}
& -FB  \frac{d}{dr}\bigg( \frac{B}{F} \frac{dF}{dr} \bigg) + \frac{B^2F^2}{r^4} = \mu ^4 + \mu ^2 \\
&- \frac{12 M  ^2}{r ^2} + \Lambda \mu ^2 r^2 + \frac{2 M \mu ^2}{r} + \frac{6M}{r}- 6 M \Lambda r + \frac{ \mu ^4 r ^2}{F} \\
 &= \mu ^4 +2  \mu ^2 +\frac{36 M^2}{F}.
 \end{split}
\ee

\noindent Identifying with Eq. (\ref{bigone}), this means

\be 
\gamma ^2 = 36M^2 \quad, \quad  \kappa=  \mu ^2 (2 + \mu ^2).
\label{gaka}
\ee

The functions $p$ and $q$ are now explicitly given thanks to Eq. (\ref{pq}) and Eq. (\ref{gaka}), which proves the isospectrality for a metric such that $B(r)$ satisfies Eq. (\ref{metric}).\\

The potentials can also be explicitly determined, from Eq. (\ref{pote}), for both perturbations. The axial perturbation are described by:

\bea
&&V^-=  \bigg( 1 + \frac{2M}{r} + \Lambda r^2 \bigg)  \bigg[ \frac{l(l+1)}{r^2} - \frac{6M}{r^3}  \bigg],
\nonumber \\ 
&& \quad \quad  \quad \quad  \quad \quad  \quad \quad  \quad \quad 
\eea

\noindent while the polar perturbations feel the potential
 
\bea
&&V^+= \frac{2 }{r^3} \bigg( 1 + \frac{2M}{r} + \Lambda r^2 \bigg) \times \nonumber \\ &&\frac{9M^3+3L^2Mr^2 +L^2(1+ L)r^3 + 9M^2r( L - \Lambda r^2)  }{(Lr+3M)^2}.\nonumber \\ 
&& \quad \quad  \quad \quad  \quad \quad  \quad \quad  \quad \quad 
\eea

\section{Phantom gauge}

In this section, we briefly discuss  the Phantom gauge. As we deal with six equations, namely Eqs. (\ref{one}-\ref{six}), and eight unknown variables, the solutions involve two arbitrary functions. This comes from the degrees of freedom associated with the rotation of the chosen tetrad. If first order infinitesimal rotations of the tetrad basis are performed, $\Psi _0$ and $\Psi _4$ are affected at the second order level while $\Psi _1$ and $\Psi _3$ are affected at the first order level (the interested reader can find a clear proof in \citep{Chandrasekhar:1985kt}, Chapter 17.(g) or through Eq. (7.79) in \citep{Hawking:1979ig}). At the linear order which is considered here, $\Psi _0$ and $\Psi _4$ are therefore gauge invariant (not affected by infinitesimal rotations), contrarily to $\Psi _1$ and $\Psi _3$. We have chosen a gauge such that

\be
\Psi _1 = \Psi _3 = 0.
\ee

The vanishing of $\Psi_1$ and $\Psi_3$ does not affect the behavior of $\Psi_0$ and $\Psi_4$. This gauge leads to the radial equations (\ref{first}) and (\ref{second}) . \\

Another meaningful choice could have been done: the so-called ``Phantom Gauge". The previous gauge was useful to separate the equations when conditions given by Eqs. (\ref{separ1}) and (\ref{separ2}) were fulfilled. However, if these conditions are not respected it is still possible to obtain two decoupled equations. Thanks to the freedom associated with the rotation of the tetrad, one can impose two additional {\it ad hoc} constraints. By applying $Br^2 \bigg( \mathcal{D}_2^{\dag}- \frac{3}{r} \bigg)$ to Eq. (\ref{one}) and $\mathcal{L}_2$ to Eq.(\ref{two}), it is possible eliminate $\Phi_0$. Indeed the condition

\bea
-Br^2 \bigg( \mathcal{D}_2^{\dag}- \frac{3}{r} \bigg) \bigg( 6 r^3 k \Psi _2 + \sqrt{2}  r [R_1] \bigg) - \nonumber \\ \mathcal{L}_2 \bigg( 6 r^3 s \Psi _2 + 2 r^2 [R_2] \bigg)=6rB' \Phi_1
\eea

\noindent gives

\be 
[Br^2 \mathcal{D}_2^{\dag} \mathcal{D}_0 -6i \omega r +  \mathcal{L}_2 \mathcal{L}_{-1}^{\dag} ] \Phi _1 =0,
\label{eqphantom}
\ee

\noindent and therefore
\be
 [Br^2 \mathcal{D}_2^{\dag} \mathcal{D}_0 -6i \omega r ] R _1 =0.
 \label{radialphantom}
\ee

The same procedure can be followed for $\Phi _3$. This gauge might have appeared to be well suited to derive isospectrality for more general metrics, that is beyond the conditions Eqs. (\ref{separ1}) and (\ref{separ2}) . The radial equation (\ref{radialphantom}) can be written in the form of Eq. (\ref{equation1}) with $Y$ defined by 

\be 
Y=rBR_1,
\ee

 \noindent as well as $P$ and $Q$ expressed by:

\be 
P= \frac{d}{dr^*} \log \bigg( \frac{r^2}{B }\bigg),
\ee

 \noindent and 
 
 \be 
 Q=\frac{B}{r^2}( 4rB'+r^2B'' +2B + \mu ^2).
 \ee

 However, in that case, it seems difficult (if not impossible) to find $p$ and $q$ so that Eq. (\ref{c2}) and  Eq. (\ref{c3}) are fulfilled. One could follow the same procedure than previously and replace Eq. (\ref{291}) and (\ref{290}) with
 
\bea
R_{PG} &=& \zeta _{PG} V + \frac{d}{dr^*} (W + 2 i \omega \zeta _{PG}),  \label{291g}\\
\gamma _{PG} &=& - \frac{r^2}{B} \bigg( \frac{d}{dr^*} ( \zeta _{PG} V ) + WV \bigg),  \label{290g}
\eea

where $\frac{r^2}{B}$ appears instead of $\frac{r^4}{B^2}$.  Then, Eq. (\ref{constte}) is replaced by

\be
\begin{split}
\frac{r^2}{B} R_{PG} \zeta _{PG} V   + \gamma _{PG}  (W+2i \omega \zeta _{PG}) = K _{PG} = \mathrm{cte}. 
\label{constteg}
 \end{split}
 \ee
 
It is however not anymore possible to require $\gamma _{PG}= \mathrm{cte}$ and $\zeta _{PG}=1$ as it has been previously done for $\gamma$ and $\zeta$. Indeed, if $\zeta _{PG}=1$, then $\gamma _{PG}=\frac{B}{r^2} \gamma$ which cannot be constant. The phantom gauge does not seem to bring any new convenient way to go ahead in this approach.

\section{Summary and conclusion}

Let us summarize the main ingredients of the calculation. The conditions (\ref{separ1},\ref{separ2}) allow to decouple equations (\ref{one}-\ref{three}) in the form of Eq. (\ref{equation1}) with functions P and Q so that Eq. (\ref{constte}) is fulfilled. These conditions lead to the Schwarzschild-(anti-)de Sitter metric (\ref{metric}). This allows to write $ -F \bigg[ \frac{d^2}{dr^{*2}} \bigg( \log F \bigg) -\frac{B}{r^4}F \bigg] $  as 
$\mathrm{cte}+\frac{\mathrm{cte'}}{F}$, yielding explicit expressions for $p$ and $q$ which show the isospectrality. \\

In this article, we tried to go a bit deeper into the original argument from Chandrasekhar so as to make it accessible to the reader who wants to apply the method to a specific spacetime structure.  We show explicitly what are the conditions to prove isospectrality in this framework. As a result, S(A)dS black holes emerge naturally as being isospectral. This  also led us to obtain the exact form of the potential for the polar and axial perturbations. \\

Isospectrality is a beautiful property which seems to be true only for very specific geometries. As far as we know, no analytical proof of isospectrality (or of the breakdown of isospectrality) as been produced yet in full generality. This article goes slightly beyond Schwarzschild and points out the difficulties one has to face when trying to extend the proof to more general spacetimes.

\bibliography{iso}

 \end{document}